\documentclass[aps,prb,twocolumn,groupedaddress,showpacs]{revtex4}
\usepackage{epsfig}
\usepackage{color}
\begin{document}

\title{Magnetic and Transport Properties of a Coupled Hubbard Bilayer
with Electron and  Hole Doping}
\author{K. Bouadim$^1$, G.G. Batrouni$^1$, F. H\'ebert$^1$, and
  R.T. Scalettar$^2$} 
\affiliation{$^1$INLN, Universit\'e de Nice-Sophia Antipolis, CNRS;
  1361 route des Lucioles, 06560 Valbonne, France}
\affiliation{$^2$Physics Department, University of California, Davis,
  California 95616, USA}

\begin{abstract}
The single band, two dimensional Hubbard Hamiltonian has been
extensively studied as a model for high temperature superconductivity.
While Quantum Monte Carlo simulations within the dynamic cluster
approximation are now providing considerable evidence for a $d$-wave
superconducting state at low temperature, such a transition remains
well out of reach of finite lattice simulations because of the ``sign
problem".  We show here that a bilayer Hubbard model, in which one
layer is electron doped and one layer is hole doped, can be studied to
lower temperatures and exhibits an interesting signal of $d$-wave
pairing.  The results of our simulations bear resemblance to a recent
report on the magnetic and superconducting properties of
Ba$_2$Ca$_3$Cu$_4$O$_8$F$_2$ which contains both electron and hole
doped CuO$_2$ planes.  We also explore the phase diagram of bilayer
models in which each sheet is at half-filling.
\end{abstract}

\maketitle

\section{Introduction} 

The single band, two dimensional Hubbard Hamiltonian provides one
possible microscopic model for pairing which is driven by electronic
correlations rather than the interactions of electrons with the
lattice.  Many analytic and numeric \cite{numericreview} treatments
suggest that there may indeed be a superconducting phase at low
temperature away from half-filling in this model.  The issue is a
difficult one, however, owing to the likely existence of a variety of
different phases which are close in energy on the one hand, and the
nature of the approximations made in the solution on the other.  Exact
diagonalization studies \cite{lin88,sorella}, while very useful, are
typically on lattices of only a few tens of sites, and hence finite
size effects are a considerable concern.  Quantum Monte Carlo (QMC)
\cite{hirsch88,white89p}, which can in principle address the issue in
an unbiased way (on lattices an order of magnitude or more larger than
diagonalization) has been unable to access sufficiently low
temperatures due to the `minus sign problem' \cite{loh90}.

Recently, progress has been made using improved numerical methods.
The `density matrix renormalization group' has pushed forward from one
dimension to address geometries of many coupled chains\cite{dmrg}.
The dynamic cluster approximation has improved on dynamical mean field
treatments by showing the robustness of a finite temperature
transition to a superconducing state as an increasingly fine momentum
grid is incorporated in the self-energy \cite{dca}.  Nevertheless,
there is still numeric work which contests the conclusion that the two
dimensional Hubbard Hamiltonian has long range $d$-wave pair
correlations\cite{imada07}.

In this paper, we present determinant Quantum Monte Carlo (DQMC)
calculations of a bilayer Hubbard model for which we are able to
attain much lower temperatures than the single layer case.
Specifically, by doping the two layers symmetrically about
half-filling, $\rho=1$, we find that the sign problem is greatly
reduced, allowing simulations at temperatures which are roughly two
orders of magnitude below the bandwidth, $T \approx W/100$.  In single
layer simulations of the doped system, the lowest attainable
temperatures are $T \approx W/40$.  Previous DQMC studies of bilayer
models have looked at the case when both layers are half-filled, and
examined magnetic order-disorder transitions which occur as the
interlayer hopping is increased \cite{scalettar94}.  A decreasing
interlayer hopping monotonically reduces the pairing correlations in
this situation.

Our work is partially motivated by studies of
Ba$_2$Ca$_3$Cu$_4$O$_8$F$_2$ and Ba$_2$Ca$_2$Cu$_4$O$_6$F$_2$ which
are an experimental realization of materials in which electron and
hole doped sheets coexist within the family of cuprate superconductors
\cite{shimizu07}.  In the former, four-layered compound, the two outer
planes are electron-doped with $N_{e} \approx 0.06 - 0.08$, while the
two inner planes are hole doped roughly symetrically, that is $N_h
\approx 0.06 - 0.08$.  The superconducting transition temperature is
$T_c = 55$ K, and pairing coexists with long range antiferromagnetic
order with N\'eel temperature $T_N = 100$ K.  The latter,
three-layered compound has outer plane doping $N_{e} \approx 0.06 -
0.08$, but a larger inner plane doping $N_h \approx 0.13$.  Its
superconducting $T_c = 76$K with only short range antiferromagnetic
correlations.  This is attributed to a decoupling of the magnetism of
the electron doped outer planes by the large doping of the inner plane
\cite{shimizu07}.

\vskip0.2in
\section{Model and Methodology}

In order to model such materials, we consider the two layer Hubbard
Hamiltonian,
\begin{eqnarray}
H&=& -t \sum_{\langle {\bf i},{\bf j} \rangle \, m\sigma} \big(
c^\dagger_{{\bf j}\,m\sigma} c_{{\bf i}\,m\sigma} + 
{\rm h.c.} \big)
\nonumber \\
&-&t_\perp \sum_{{\bf i} \sigma}
\big( c^\dagger_{{\bf i}\,1\sigma} c_{{\bf i}\,2\sigma} + 
{\rm h.c.} \big)
- \sum_{{\bf i}\, m \sigma} \mu_m n_{{\bf i}\, m\sigma}
\nonumber \\
&+& U \sum_{\bf{i}\, m} 
(n_{{\bf i}\, m\uparrow}-\frac12) (n_{{\bf i}\,m \downarrow}-\frac12) \,\,. 
\label{Hamilton}
\end{eqnarray}
The first term is the usual hopping of electrons between near neighbor
sites ${\bf i}$ and ${\bf j}$ of a two dimensional square lattice.
Unless otherwise stated, the results in this paper are for two coupled
8x8 lattices. The electrons in the kinetic energy term have a spin
index $\sigma = \uparrow,\downarrow$ and also a layer index $m=1,2$.
The second term is an interlayer hopping.  The third term is a
layer-dependent chemical potential.  We will choose $\mu_1 = - \mu_2$
to produce layers which have opposite dopings.  Finally, electrons of
opposite spin on the same site of the same layer feel a repulsion $U$.

Our simulations employ the DQMC algorithm
\cite{blankenbecler81,white89p} in which a path integral is written
for the partition function, the fermion interactions are replaced by a
coupling to an auxiliary Hubbard-Stratonovich field, and then the
fermion degrees of freedom are integrated out analytically.  The
method produces exact results on the lattice sizes considered, apart
from `Trotter' errors associated with the imaginary time
discretization, which we have verified are smaller than our
statistical error bars.

The magnetic properties are determined from the spin-spin
correlations,
\begin{eqnarray}
c({\bf l}) = \langle \, M_{{\bf j}+{\bf l},m}^z
M_{{\bf j},m}^z \, \rangle
\nonumber \\
M_{{\bf j},m}^z = 
n_{{\bf j} m \uparrow} - n_{{\bf j}  m \downarrow}
\,\, ,
\end{eqnarray}
which are independent of layer index $m$ because of our choice of
symmetric doping and the particle-hole symmetry of the Hubbard
Hamiltonian.  The Fourier transform gives the structure factor,
\begin{eqnarray}
S({\bf q}) = \sum_{{\bf l}} e^{i {\bf q} \cdot {\bf l}} c({\bf l})\,\,.
\end{eqnarray}
At half-filling, $S({\bf q})$ is largest at the antiferromagnetic
wavevector ${\bf q} = (\pi,\pi,\pi)$.

A first insight into the metal-insulator transition can be obtained
from the zero momentum spectral function (density of states)
$A(\omega)$ which is determined from the Greens function,
\begin{eqnarray}
G_{{\bf i-j}, \, m \sigma}(\tau) = \langle c_{{\bf i} \, m \sigma}(\tau) 
c_{{\bf j} \, m \sigma}^\dagger(0) \rangle
\nonumber \\
G_{{\bf k} \, m \sigma}(\tau) = \sum_{\bf l} e^{ i {\bf k} \cdot {\bf l} }
\,\, G_{{\bf l} \, m \sigma}(\tau) 
\nonumber \\
A(\omega) = \int_0^\beta \, d\tau \, 
\frac{e^{-\omega \tau}}{e^{\beta \omega} +1}
\,\, \sum_{m\sigma} G_{{\bf k=0}, \, m \sigma}(\tau)
\end{eqnarray}
using the maximum entropy method \cite{gubernatis}.

The dc conductivity $\sigma_{\rm dc}$ 
also characterizes the metal-insulator transition,
and is measured from the current-current correlation function,
\begin{eqnarray}
j_x({\bf l},\tau)
&=& e^{H \tau} \, j_x({\bf l},0) e^{-H \tau}
\nonumber
\\
j_x({\bf l},0)
&=&  i t \sum_{m \sigma} 
( c^\dagger_{{\bf l}+x \, m \sigma}
c_{{\bf l}\, m \sigma}
- c^\dagger_{{\bf l}\, m \sigma}
c_{{\bf l}+x \, m \sigma} ) 
\nonumber
\\
\Lambda_{xx}({\bf q}; \tau)
&=& \sum_{\bf l} e^{i {\bf q} \cdot {\bf l} } 
\langle j_x({\bf l},\tau) j_x(0,0) \rangle
\end{eqnarray}
This imaginary time quantity, which comes directly out of the
determinant QMC simulations, is related to the real frequency response
by the fluctuation-dissipation theorem,
\begin{eqnarray}
\Lambda_{xx}({\bf q}; \tau)
= \int_{-\infty}^{+\infty} \, \frac{d\omega}{\pi} \, 
\frac{e^{-\omega \tau}}{1-e^{-\beta \omega}}
\, {\rm Im} \Lambda_{xx}({\bf q},\omega).
\end{eqnarray}
As discussed in [\onlinecite{trivedi96}], at sufficiently low
temperatures we can replace ${\rm Im}\Lambda$ by its low frequency
behavior ${\rm Im}\Lambda \approx \sigma_{\rm dc} \omega$, leading to
the relation,
\begin{eqnarray}
\Lambda_{xx}({\bf q}=0; \tau = \frac{\beta}{2})
= \frac{\pi \sigma_{\rm dc}}{\beta^2}  \,\,.
\end{eqnarray}
This enables us to obtain the conductivity directly from the imaginary
time response without the necessity for analytic continuation, which
is more difficult for two particle response functions, like the
current-current correlator, than for the single particle Greens
function, owing to their larger fluctuations.

To describe superconductivity, we compute the correlated pair field
susceptibility, $P_{\alpha}$, in different symmetry channels,
\begin{eqnarray}
P_{\alpha} &=& \int_0^\beta d \tau \langle \Delta_{\alpha}(\tau)
\Delta_{\alpha}^\dagger(0) \rangle \nonumber \\
\Delta^\dagger_{\alpha} &=& {1 \over N} \sum_{\bf k} f_\alpha ({\bf
k}) c^\dagger_{{\bf k}\uparrow} c^\dagger_{-{\bf k}\downarrow}
\nonumber \\ f_s({\bf k}) &=& 1 \nonumber \\ f_{s^*}({\bf k}) &=& {\rm
cos} \, k_x + {\rm cos} \, k_y \nonumber \\ f_d({\bf k}) &=& {\rm cos}
\, k_x - {\rm cos} \, k_y \,\,.
\label{Pmom}
\end{eqnarray}

The correlated susceptibility $P_\alpha$ takes the expectation value
of the product of the four fermion operators entering Eq.~\ref{Pmom}.
We also define the uncorrelated pair field susceptibility
$\overline{P}_\alpha$ which instead computes the expectation values of
pairs of operators {\it prior} to taking the product.  Thus, for
example, in the $s$-wave channel,
\begin{eqnarray}
P_{s} &=& 
{1 \over N^2} 
\sum_{{\bf i},{\bf j}}
\int_0^\beta d \tau 
\langle \,
c_{{\bf i}\downarrow}(\tau) \, c_{{\bf i}\uparrow}(\tau) 
\, c^\dagger_{{\bf j}\uparrow}(0) \, c^\dagger_{{\bf j}\downarrow}(0) 
\, \rangle 
\nonumber \\
\overline{P}_{s} &=& 
{1 \over N^2} 
\sum_{{\bf i},{\bf j}}
\int_0^\beta d \tau 
\langle \,
c_{{\bf i}\downarrow}(\tau) 
\, c^\dagger_{{\bf j}\downarrow}(0) 
\, \rangle  \,\,
\langle \,
c_{{\bf i}\uparrow}(\tau) 
\, c^\dagger_{{\bf j}\uparrow}(0) 
\, \rangle 
\,\,.
\label{Pbar}
\end{eqnarray}
$P_\alpha$ includes both the renormalization of the propagation of the
individual fermions as well as the interaction vertex between them,
whereas $\overline{P}_\alpha$ includes only the former effect.  Indeed
by evaluating both $P$ and $\overline{P}$ we are able to extract
\cite{white89} the interaction vertex $\Gamma$,
\begin{eqnarray}
\Gamma_\alpha = {1 \over P_\alpha}
- {1 \over \overline{P}_\alpha} \,\,.
\end{eqnarray}
If $\Gamma_\alpha \bar P_\alpha < 0$, the associated pairing
interaction is attractive.  In fact, rewriting Eq.~10 as,
\begin{eqnarray}
P_\alpha = \frac{\bar P_\alpha}{1+\Gamma_\alpha \bar P_\alpha}
\end{eqnarray}
suggests that $\Gamma_\alpha \bar P_\alpha \rightarrow -1$ signals a
superconducting instability.  We will discuss this criterion in more
detail in the coming sections.

\vskip0.2in
\section{Bilayer Phase Diagram at Half-Filling}

\begin{figure}
\centerline{\epsfig{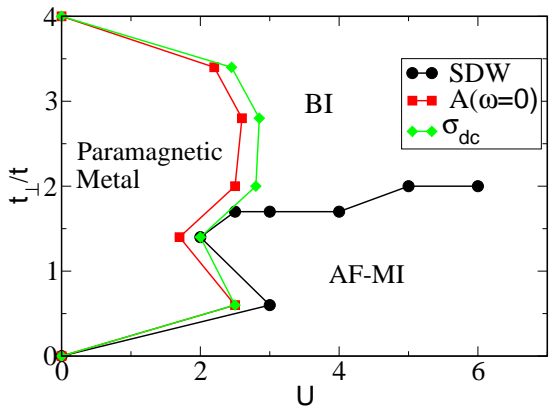}}
\caption{(Color online) Phase diagram for the half-filled bilayer
Hubbard model.  A paramagnetic metallic phase is present at weak
coupling.  At large coupling there is a transition from an
antiferromagnetic Mott-insulating phase to a paramagnetic
band-insulating phase.  The phase boundaries obtained by the
conductivity $\sigma$ and density of states at the Fermi level,
$A(0)$, are consistent.  }
\label{Phasediagram}
\end{figure}

We begin with the phase diagram at half-filling, that is when $\mu_1 =
\mu_2 =0$, and both layers have equal occupation $\rho_1 = \rho_2 =1$.
(Note there is no sign problem in this case because of particle-hole
symmetry.) Here we do not expect superconductivity.  Nevertheless
there is an interesting competition between Mott insulating behavior
when $U$ is the dominant energy scale, and band insulating behavior
for large $t_\perp$.  Indeed, increased interlayer coupling suppresses
the antiferromagnetic correlations which are present in the Mott
phase, since $t_\perp$ promotes the formation of interlayer singlets
between the two spatial sites immediately above and below each other.
These spin-0 singlets are magnetically decoupled, destroying long
range spin order.  Earlier determinant QMC studies determined the
critical value of $t_\perp \approx 1.6$ for this AF-PM transition
\cite{scalettar94}.

The strong coupling region of Fig.~\ref{Phasediagram} exhibits this
phenomena, and yields a $(t_\perp/t)_c$ consistent with the earlier
study \cite{scalettar94}.  At weak coupling, however, this
insulator-insulator transition is replaced by a metallic phase.
Previous cluster DMFT \cite{kancharla07} studies of the bilayer model
show a phase diagram which is in qualitative agreement with
Fig.~\ref{Phasediagram}.  We will compare the results of the two
methods in more detail at the end of this section.  First, we will
describe in detail how this phase diagram is obtained.

\begin{figure}
\centerline{\epsfig{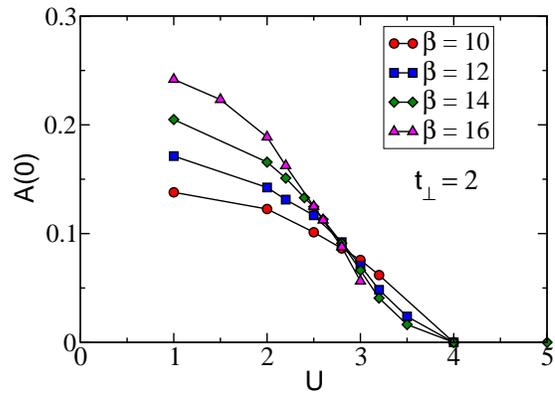}}
\caption{(Color online) Density of states at the Fermi surface $A(0)$
for $t_\perp=xx$.
At weak $U$, $A(0)$ rises as $T=1/\beta$ is lowered, indicating a
metallic phase with nonzero Fermi level density of states.  In
contrast, at large $U$, $A(0)$ falls with decreasing $T$, indicating
insulating behavior.  $(U/t)_c \approx 2.8$.  }
\label{A0vsUN8tp2.0}
\end{figure}

In Fig.~\ref{A0vsUN8tp2.0} the density of states at the Fermi surface,
$A(\omega=0)$ is shown for four temperatures along a horizontal cut
through the phase diagram at fixed $t_\perp/t=2$.  At weak coupling,
the low temperature limit is non-zero, indicating a metallic phase,
while at strong coupling, $A(\omega=0)$ decreases as $T$ is lowered.
We conclude that at the crossing point $U/t \approx 2.8$ a
metal-insulator transition occurs.

\begin{figure}
\centerline{\epsfig{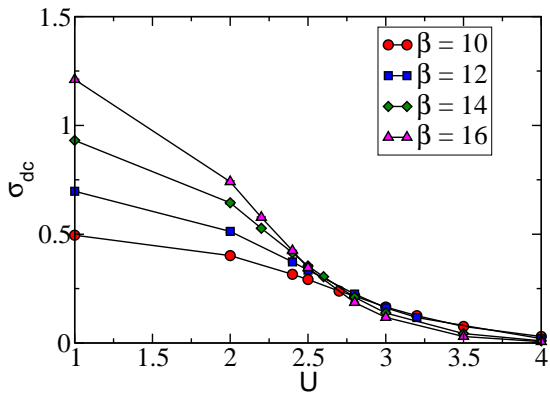}}
\caption{(Color online) The conductivity $\sigma_{\rm dc}$ for a
horizontal cut (fixed $t_\perp/t=3.4$) and varying $U/t$ through the
phase diagram.  Values at four inverse temperatures are given.  As
with the density of states at the Fermi energy, $A(\omega=0)$, shown
in Fig.~\ref{A0vsUN8tp2.0}, the conductivity exhibits a crossing
pattern which gives the location of the metal-insulator phase
boundary: $\sigma_{\rm dc}$ increases as $\beta$ increases (metallic
behavior) below $U/t \approx 2.6$, and falls as $\beta$ increases
above this value.  }
\label{SvsUtp3.4N8}
\end{figure}

In Fig.~\ref{SvsUtp3.4N8} we see that the conductivity $\sigma_{\rm
dc}$ similarly can determine the location of the metal-insulator phase
boundary.  Here a change in the temperature behavior of the
conductivity, from increasing as $T$ is lowered (metallic) to
decreasing when $T$ is lowered (insulating) occurs at $U/t \approx
2.6$ when the interlayer hopping is $t_\perp/t =3.4$.

Multiple horizontal (constant $t_\perp/t$) cuts through the phase
diagram similar to those of Figs.~\ref{A0vsUN8tp2.0}-\ref{SvsUtp3.4N8}
were used to generate the metal-insulator phase boundary of
Fig.~\ref{Phasediagram}.  Note the consistency of the locations of the
critical interaction strengths between those obtained from the density
of states $A(\omega=0)$ (red squares in Fig.~\ref{Phasediagram}) and
the conductivity $\sigma_{\rm dc}$ (green diamonds in
Fig.~\ref{Phasediagram}).

\begin{figure}
\centerline{\epsfig{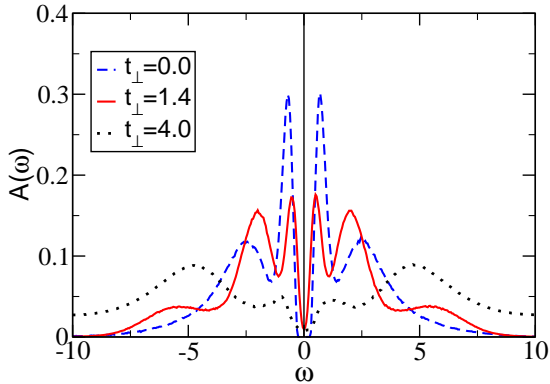}}
\caption{(Color online) Density of states $A(\omega)$ at $U/t=3$ and
inverse temperature $\beta=14$, showing insulating behavior at all
values of interlayer coupling.  $t_\perp=0$ and $t_\perp=1.4$ are
Mott/Slater insulating phases with a gap produced by a combination of
the on-site repulsion and antiferromagnetic spin correlations.
$t_\perp=4.0$ has a gap which is primarily band insulator in
character.  }
\label{AomegaU3N8B14}
\end{figure}

In this bilayer model, at half-filling $\mu_1=\mu_2=0$, the
suppression of the zero frequency spectral weight can come from any of
three mechanisms: the opening of a band gap at sufficiently large
$t_\perp$, a ``Slater gap'' created by antiferromagnetic fluctuations
which can form on a scale set by the exchange constant $J \propto
t^2/U$, and a ``Mott gap'' between the upper and lower Hubbard bands
when $U$ exceeds the bandwidth $W$.  (The bandwidth $W= 8t$ at
$t_\perp=0$.) In general, these different insulating phases cross over
to each other more or less smoothly, although the Slater insulator can
be distinguished by the presence of long range spin correlations.
Fig.~\ref{AomegaU3N8B14} shows the full frequency dependence of the
density of states at $U/t=3$ and three values of $t_\perp$, all of
which exhibit a gap in $A(\omega)$.  (The non-zero residual values of
$A(\omega)$ for $t_\perp=1.4$ and $4.0$ will be driven to zero if
$\beta$ is increased.  See Fig.~\ref{A0vsUN8tp2.0}.) From
Fig.~\ref{AomegaU3N8B14} we infer that the phase diagram is insulating
all along the vertical line $U/t=3$ in Fig.~\ref{Phasediagram}.

\begin{figure}
\centerline{\epsfig{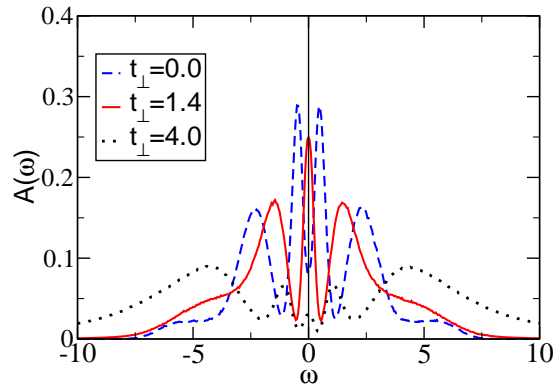}}
\caption{(Color online) The same as Fig.~\ref{AomegaU3N8B14}, except
$U/t=2$.  Although $t_\perp=0.0$ and $4.0$ are still insulating, the
density of states for $t_\perp=1.4$ has a peak at $\omega=0$ and is
metallic in character.  }
\label{AomegaU2N8B14}
\end{figure}

In contrast, Fig.~\ref{AomegaU2N8B14}, which shows the same three
values of $t_\perp$ except at weaker coupling, $U/t=2$, clearly
exhibits metallic behavior for the intermediate value of the
interlayer hopping.  This is one indication of the outward extent of
the metallic region from $U/t=0$ in the phase diagram of
Fig.~\ref{Phasediagram}.

\begin{figure}
\centerline{\epsfig{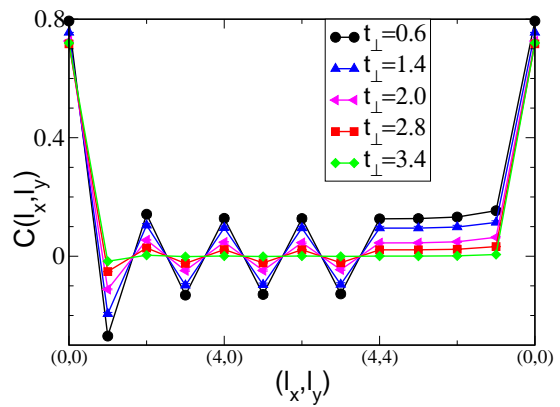}}
\caption{(Color online) Real space spin correlations at $U/t=5$.  As
$t_\perp$ increases, the antiferromagnetism is suppressed.  The
inverse temperature $\beta=14$.  }
\label{ZZN8B14U5.0HalfFillingP}
\end{figure}

\begin{figure}
\centerline{\epsfig{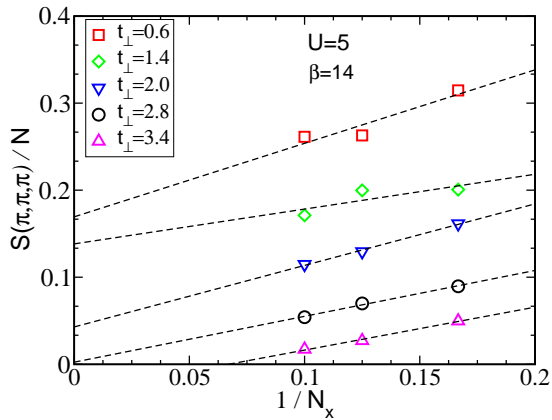}}
\caption{(Color online) Scaling of the antiferromagnetic structure
factor at $U=xx$.  If there are long range correlations,
$S(\pi,\pi,\pi)$ should grow linearly with lattice size $N$, so that
$S(\pi,\pi,\pi)/N$ approaches a constant for large $N$.  Spin wave
theory predicts a $1/N_x$ correction, where $N_x$ is the linear
lattice size ($N_x^2 = N)$.  Here we see long range order for the
three smallest values $t_\perp/t=0.4, 1.4, 2.0$, but not for the two
largest values $t_\perp/t=2.8, 3.4$.  }
\label{AF-finiteB14U5.0P}
\end{figure}

\begin{figure}
\centerline{\epsfig{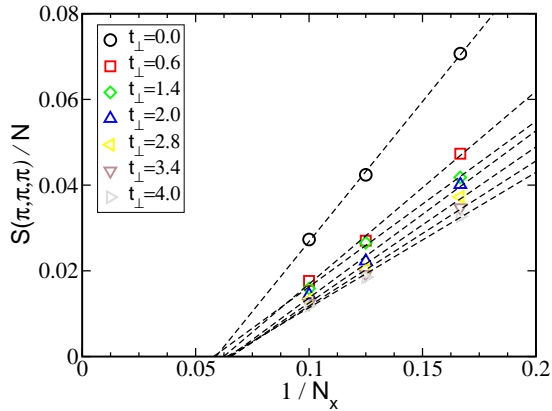}}
\caption{(Color online) Same as Fig.~\ref{AF-finiteB14U5.0P} except
$U/t=2$.  }
\label{AF-finiteU2.0P}
\end{figure}

We turn now to the spin correlations.
Fig.~\ref{ZZN8B14U5.0HalfFillingP} shows the real space spin
correlations for $U/t=5$ and different interlayer hoppings.  $t_\perp$
drives the formation of interlayer singlets which interfere with the
magnetic order.  A finite size scaling analysis is shown in
Fig.~\ref{AF-finiteB14U5.0P} where the structure factor is plotted as
a function of the inverse linear system size.  Spin wave theory
predicts \cite{huse88} that the finite size corrections to
$S(\pi,\pi,\pi)$ should be linear in $1/N_x$, with the $N_x
\rightarrow \infty$ intercept proportional to the square of the order
parameter.  We see that the order parameter is non-zero for
$t_\perp/t=0.6$ and $1.4$ and is zero for $t_\perp/t=2.8$ and $3.4$.
Somewhere in the vicinity of $t_\perp/t \approx 2$, the long range
magnetic order is destroyed.  Fig.~\ref{AF-finiteU2.0P} shows a
similar finite size scaling analysis for weaker coupling, $U/t=2$.
There is no long range magnetic order for any value of $t_\perp/t$.

Multiple vertical (constant $U/t$) cuts through the phase diagram
similar to those of
Figs.~\ref{ZZN8B14U5.0HalfFillingP}-\ref{AF-finiteU2.0P} were used to
generate the limit of the antiferromagnetically ordered regions of the
phase diagram Fig.~\ref{Phasediagram}.  This value is consistent with
previous DQMC studies \cite{scalettar94} and cluster DMFT
\cite{kancharla07}.

We conclude this section with a more quantitative comparison of
Fig.~\ref{Phasediagram} with the results obtained in cluster DMFT
\cite{kancharla07}.  At strong coupling, the AF insulator to
paramagnetic (bond) insulator transition is found by both methods to
have the same value $t_{\perp}/t = 2$.  Likewise, in both approaches,
the base of the metallic phase at $U=0$ extends from $t_{\perp}/t=0$
to $t_{\perp}/t=4$, as indeed it must analytically from the
non-interacting dispersion which has bonding and anti-bonding bands,
\begin{eqnarray}
\epsilon_1({\bf k}) &=& -t_\perp + 2t \,(\, {\rm cos} \, k_x
+ {\rm cos} \, k_y \, )
\nonumber
\\
\epsilon_2({\bf k}) &=& +t_\perp + 2t \,(\, {\rm cos} \, k_x
+ {\rm cos} \, k_y \, )  \,\,\,.
\end{eqnarray}

The extent of the metallic phase as $U$ increases from the
non-interacting limit differs quantitatively in the two methods.  The
DQMC results reported here indicate an upper limit of $U/t \approx 3$,
while within cluster DMFT the metallic region extends out to $U/t
\approx 8$.  The precise origin of this disagreement is not clear.
The peak of the cluster DMFT metallic lobe follows the emerging
AF-band insulator line rather narrowly, and it is possible DQMC cannot
resolve this small region adequately.  While the results of Figs.
\ref{A0vsUN8tp2.0} and \ref{AomegaU3N8B14} seem unambiguously to rule
out metallic behavior much beyond $U/t \approx 3$, they are on
lattices of finite extent ($N$=8x8).  Cluster DMFT works in the
thermodynamic limit and hence typically produces sharper transitions
which can distinguish narrow regions of phase space.  On the other
hand, DQMC incorporates the full momentum dependence of the
self-energy, in contrast to the 2x2 momentum grid used in cluster
DMFT.

\section{Superconductivity in the Doped System}

\begin{figure}
\centerline{\epsfig{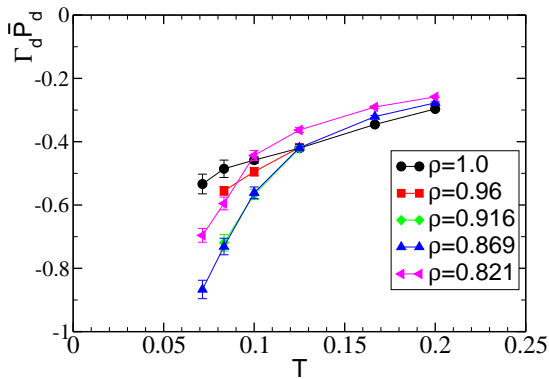}}
\caption{(Color online) $d$-wave pairing vertex as a function of
temperature for two 8x8 bilayers with interlayer hopping $t_\perp=0.6
t$.  The on-site interaction $U=3 t$.  Three fillings are shown.  Note
the close approach to $\Gamma \bar P=-1$, the onset point of a pairing
instability, and the non-monotonic dependence on filling.  The
greatest tendency to pairing is at $\rho \approx 0.87$.  }
\label{gammadvsT}
\end{figure}

Fig.~\ref{gammadvsT} shows a central result of our paper.  The product
of the $d$-wave pairing vertex and the uncorrelated susceptibility,
$\Gamma_d \bar P_d$, is seen to turn sharply negative (attractive) as
the temperature $T$ is lowered.  As described in Eq.~11, $\Gamma_d
\bar P_d \rightarrow -1$ in principle would signal a superconducting
instability.  For $\rho \approx 0.87$, $\Gamma_d \bar P_d \approx
-0.9$ .  In comparison, the most negative $\Gamma_d \bar P_d$ reported
\cite{white89} for the single band model is $\Gamma_d \bar P_d = -0.45
$ at half-filling, $\rho=1.000$, and $\Gamma_d \bar P_d = -0.25 $ for
doping to $\rho = 0.875$.  It should be kept in mind, however, that
the lowest accessible temperature in the latter case is $\beta = 6/t$.
At the same $\beta= 6/t$ and doping $\rho=0.875$, as seen in
Fig.~\ref{gammadvsT}, the bilayer system has a somewhat more negative
$\Gamma_d \bar P_d = -0.31$.  Thus the approach of $\Gamma_d \bar P_d$
to $-1$ seen in the bilayer system is due both to a more attractive
pairing vertex, but also due to the ability of the simulation to reach
much colder temperatures.

\begin{figure}
\centerline{\epsfig{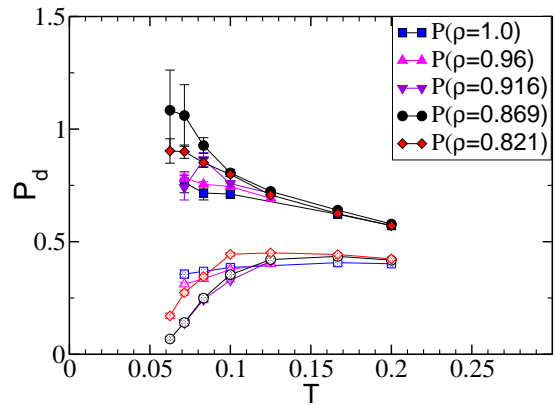}}
\caption{(Color online) Correlated (closed symbols) and uncorrelated
(open symbols) $d$-wave pairing susceptibility as a function of
temperature for two 8x8 bilayers with interlayer hopping $t_\perp=0.6
t$.  The on-site interaction $U=3 t$.  Five fillings are shown.  In
all cases the vertex is attractive, ie.~$P_d > \bar P_d$.  The degree
of attraction is non-monotonic, first increasing with doping, but then
declining.  }
\label{PdvsT}
\end{figure}

Although we find the vertex $\Gamma_d \bar P_d$ approaches -1, this
criterion for an instability is incomplete.  One also needs to require
that the uncorrelated susceptibility $\bar P$ remain finite at the
transition point.  Especially in the situation where there is
competing order (e.g. antiferromagnetism and pairing) it is possible
for the uncorrelated susceptibility of one type of order to be driven
to small values by the other order, so that even though the vertex
approaches -1, order in this channel is usurped.  Fig.~\ref{PdvsT}
addresses this issue for the bilayer model.  Despite the fact that
$\Gamma_d \bar P_d$ is getting close to $-1$, the correlated vertex
$P_d$ grows relatively slowly as $T$ is decreased.  The reason is
clear from Fig.~\ref{PdvsT} in which it is seen that the uncorrelated
susceptibility is rapidly dropping as $T$ is lowered.  This is rather
different from the doped single layer model, where $\bar P_d$ grows as
$T$ is lowered.  (At half-filling in the single layer model $\bar P_d$
declines slightly as $T$ is decreased, as found here also in the
bilayer model.)

\begin{figure}
\centerline{\epsfig{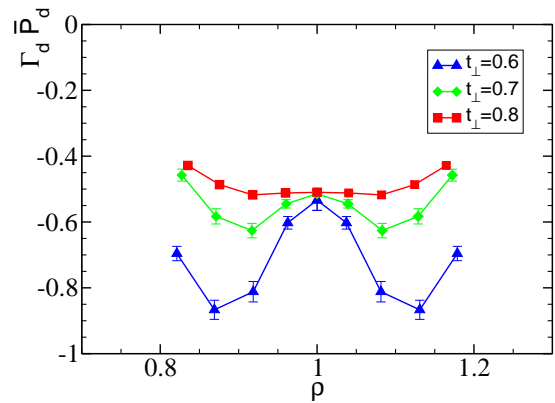}}
\caption{(Color online) $d$-wave pairing vertex as a function of
filling for two 8x8 bilayers with interlayer hopping $t_\perp/t=0.6,
0.7, 0.8$.  The on-site interaction $U=3 t$ and inverse temperature
$\beta=14$.  The greatest tendency to pairing is at $\rho \approx
0.87$.  }
\label{gammadvsrho}
\end{figure}

An interesting feature of Figs.~\ref{gammadvsT} and \ref{PdvsT} is
that the $d$-wave attraction is maximal at $\rho \approx 0.87$,
whether measured via the vertex or the correlated susceptibility.
This point is made more concretely in Fig.~\ref{gammadvsrho}.  The
behavior of the $d$-wave superconducting vertex bears an interesting
resemblance to the superconducting ``domes" of the cuprate materials
in which the transition temperatures are maximized a finite distance
away from ``half-filling" (one hole per Cu).  Indeed, even the values
of the doping which maximizes $T_c$ and the width of the base of the
dome are in reasonable quantitative agreement.  It is to be emphasized
that, within the same DQMC methodology, the single layer Hubbard model
has a maximum pairing vertex at half-filling.
Fig.~\ref{gammadvsrho} also indicates that, within the parameter range
accessible, the degree of enhancement increases as $t_\perp$
decreases.  Eventually we expect this trend to reverse, since at
$t_\perp=0$, the single layer model, there is a lesser tendency for
pairing.  (We cannot accumulate data for smaller values of $t_\perp$
because the sign problem prevents simulations at as low a temperature
as for the data shown.)

\begin{figure}
\centerline{\epsfig{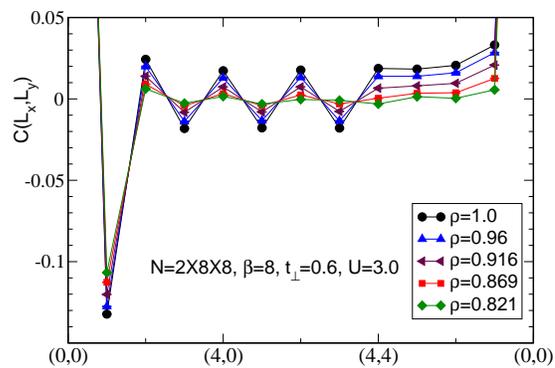}}
\caption{(Color online) Real space spin correlations.  At half
filling, $\rho=1.00$, and for small dopings, $\rho=0.96$, there is a
strong oscillatory pattern indicative of long range magnetic order
\cite{footnote2}.  For larger dopings, the spin correlations are
sharply curtailed.  }
\label{realspacespincorr}
\end{figure}

We turn now to the magnetic properties of the doped system and in
particular their connection to those observed in the cuprate
superconductors.  Fig.~\ref{realspacespincorr} shows the real space
spin-spin correlations for $\rho = 1.00, 0.96, 0.92, 0.87$ and $0.82$
at $\beta=8/t$, $U=3 t$, $T_\perp=0.6$.  These results have a
quantitative similarity to the Ba$_2$Ca$_3$Cu$_4$O$_8$F$_2$ and
Ba$_2$Ca$_2$Cu$_4$O$_6$F$_2$ materials in that the robust magnetic
correlations present for $\rho=1.00$ and $\rho=0.96$ are dramatically
suppressed for $\rho = 0.87$.  A finite Neel temperature $T_N$ is
present for the four layer compound Ba$_2$Ca$_2$Cu$_4$O$_6$F$_2$ which
has electron and hole dopings $N_e,N_h \approx 0.06$ and absent for
the three layer compound Ba$_2$Ca$_2$Cu$_4$O$_6$F$_2$ which has hole
doping $N_h \approx 0.14$ in the central layer.

Why is the sign problem ameliorated in these bilayer simulations?  In
DQMC for the single layer Hamiltonian, the operator $n_{i\uparrow}$
couples to the Hubbard Stratonovich field $h_i$ \cite{footnote}
shifted by the chemical potential $h_i -\mu$.  Meanwhile, $n_{i
\downarrow}$ couples to $-h_i -\mu$.  At half-filling, $\mu=0$,
particle-hole symmetry is reflected in the fact that the up and down
species couple to the quantities $\pm h_i$ which are symmetric about
zero.  The up and down determinants can be shown to have the same
sign, and hence their product is positive.  For $\mu \neq 0$ this
symmetry and the associated connection between the signs of the two
determinants is broken, and a sign problem results.  (Note that for
the attractive Hubbard Hamiltonian $n_{i \uparrow}$ and $n_{i
\downarrow}$ both couple to $h_i -\mu$ and the two determinants are
equal at all fillings.)

Consider now the bilayer system.  We have a Hubbard-Stratonovich field
for each layer. The operators $n_{i 1 \uparrow}$ couple to $h_{i1} -
\mu$, while $n_{i 1 \downarrow}$ couple to $-h_{i1} - \mu$, and $n_{i
2 \uparrow}$ couple to $h_{i2} + \mu$, and finally $n_{i 2
\downarrow}$ couple to $-h_{i2} + \mu$, where we have explicitly set
$\mu_1 = - \mu_2 = -\mu$.  What we observe is that, to the extent that
the Hubbard-Stratonovich variables on the two layers are equal,
$n_{i1\uparrow}$ and $n_{i2\downarrow}$ are symmetrically coupled
about zero.  It is possible that this tends to lead to a positive
determinant product similar to the single layer case at half-filling.
Of course, there is no constraint that $h_{i 1} = h_{i 2}$, but we
suspect that they are nevertheless sufficiently correlated to reduce
the sign problem.

\section{Conclusions} 

In this paper we have used DQMC simulations to determine the phase
diagram, in the $(t_{\perp}/t,U)$ plane, of the half-filled bilayer
Hubbard model. Our phase diagram exhibits metallic, band insulating
and Mott insulating phases in qualitative agreement with CDMFT
results\cite{kancharla07}. However, the entire metallic phase we find
is paramagnetic with no antiferromagnetic metallic regions.

In addition, we have shown that the doped bilayer Hubbard Hamiltonian
has an attractive $d$-wave pairing vertex which approaches close to
$\Gamma_d \bar P_d = -1$, signaling a superconducting transition.
This value is much more singular than that observed in the single
layer model, partly because it is more attractive when compared at the
same inverse temperature, and partly because it is possible to
simulate to values of $\beta$ which are two to three times larger than
for a single plane.  However, the uncorrelated $\bar P_d$ gets small,
so that the enhancement of the correlated $P_d$ is not very dramatic.
On the other hand, and unlike what happens in the single
layer $d=2$ Hubbard model, the enhancement here is maximum when the
system is doped, in agreement with the phenomenology of cuprate
superconductors.

Pairing in systems with separate electron and hole doped sheets has a
long history in the context of exciton condensation
\cite{balatsky04walsh06}, but our primary motivation here has been the
recent report of cuprate-based systems Ba$_2$Ca$_3$Cu$_4$O$_8$F$_2$
and Ba$_2$Ca$_2$Cu$_4$O$_6$F$_2$ which have both types of dopings
\cite{shimizu07}.  Our results for the magnetic and pairing
correlations bear interesting connections to those materials.  While
the bilayer simulations reported here contain the the essential
feature of coupled electron and hole doped layers, it is natural to
consider direct numerics of three and four layer compounds.  Such
studies will require an order of magnitude greater simulation time,
and also have an at present unknown sign problem.  In general the sign
problem becomes worse with lattice size (and hence with number of
layers), but it is tempting to speculate that, if the above picture of
correlated determinant signs presented above is correct, similar
correlations might provide protection for larger numbers of layers.

KB, FH, and GGB acknowledge financial support from a grant from the
CNRS (France) PICS 18796, RTS from DOE-DE-FC0206ER25793.  We
acknowledge very useful help from M. Schram and K. Dawson.  This
research was supported in part by the National Science Foundation
under Grant No.~NSF PHY05-51164, and has been assigned KITP preprint
number NSF-KITP-07-195.


\begin{thebibliography}{40}
\bibitem{footnote2}
Actually, it is likely the model exhibits long range order only
at $\rho=1$.  Nevertheless, the correlations at $\rho=0.96$ decay
very little with increasing separation.  In a real material, even weak
three dimensional coupling is likely to stabilize the magnetism.
\bibitem{footnote}
Actually, the Hubbard-Stratonovich field also depends on 
imaginary time $\tau$, a dependence we suppress in this discussion.
\end{thebibliography}
\end{document}